\documentclass[12pt]{article}
\usepackage[letterpaper, margin=1in, tmargin=1in, bmargin=1in]{geometry}

\usepackage{amsthm}
\usepackage{amsmath}
\usepackage{amsfonts}
\usepackage{amssymb}
\usepackage{graphicx}
\usepackage{setspace}
\usepackage{bbm}
\usepackage{indentfirst}
\usepackage{caption}
\usepackage{subcaption}

\usepackage{array}
\usepackage{mathtools}
\usepackage[dvipsnames]{xcolor}
\usepackage{booktabs,caption}
\usepackage[flushleft]{threeparttable}
\usepackage{placeins}
\usepackage{rotating}
\usepackage{pdflscape}

\usepackage{tabularray}
\usepackage{float}
\usepackage{graphicx}
\usepackage{codehigh}
\usepackage[normalem]{ulem}

\UseTblrLibrary{booktabs}
\UseTblrLibrary{siunitx}

\NewTableCommand{\tinytableDefineColor}[3]{\definecolor{#1}{#2}{#3}}

\usepackage{tikz}
\usetikzlibrary{bayesnet}
\usetikzlibrary{arrows}
\usepackage{pgflibraryarrows}
\usepackage{pgfplots}
\usetikzlibrary{patterns,intersections,arrows.meta}

\usepackage{har2nat}  
\setcitestyle{aysep={,}}

\usepackage{hyperref}
\hypersetup{
    colorlinks=true,
    linkcolor=black,
    filecolor=magenta,      
    urlcolor=RoyalBlue,
    citecolor=RoyalBlue
}

\newcommand{\beginsupplement}{%
        \onecolumn
        \setcounter{table}{0}
        \renewcommand{\thetable}{S\arabic{table}}%
        \setcounter{figure}{0}
        \renewcommand{\thefigure}{S\arabic{figure}}%
        \setcounter{section}{0}
        \renewcommand{\thesection}{S\arabic{section}}%
     }

\begin{document}

\title{A light-touch AI literacy intervention helps protect against AI political persuasion}

\author{Reed Orchinik\thanks{Department of Social and Decision Sciences, Carnegie Mellon University, Pittsburgh, PA 15213. orchinik@cmu.edu. Corresponding author.} \and David Rand\thanks{Department of Information Science and Department of Marketing and Management Communications, Cornell University, Ithaca, NY 14850. dgr7@cornell.edu}}

\maketitle

\begin{abstract}
Conversations with large language models (LLMs) can substantially shift beliefs and attitudes, raising concerns about manipulation using AI persuasion. Here we test whether a light-touch AI literacy intervention -- a brief warning that LLMs can be prompted to persuade and may present information selectively -- helps protect users. Across two experiments (total N = 3,208 Americans) in which participants conversed with an LLM instructed to shift their views about different political topics, the presence of a warning reduced belief change by roughly one-half (-48.1\%, 95\% CI [-59.5\%, -36.8\%]) relative to the control. Importantly, the warning did not significantly reduce trust in generative AI more broadly. Light-touch literacy interventions can help protect users against AI political persuasion.


\large{\textbf{\textcolor{red}{Note: This paper has not yet undergone peer review.}}}
\end{abstract}

\newpage
\doublespacing
\section{Introduction}\label{sec:intro}
There is increasing evidence that conversations with AI chatbots can be highly persuasive. While these conversations can be used in beneficial ways, such as debunking conspiracy theories \cite{costello_durably_2024} or reducing science skepticism \cite{hornsey_using_2026}, there is substantial concern about AI dialogues being used to persuade in a harmful manner, such as persuading voters about political issues \cite{hackenburg_levers_2025, salvi_conversational_2025, argyle_testing_2025} and candidates \cite{lin_persuading_2025, potter_hidden_2024}, and eroding democratic norms \cite{schroeder_how_2025}.

\begin{figure}[t]
  \centering
  \begin{subfigure}[b]{0.48\textwidth}
    \centering
    \includegraphics[width=\linewidth]{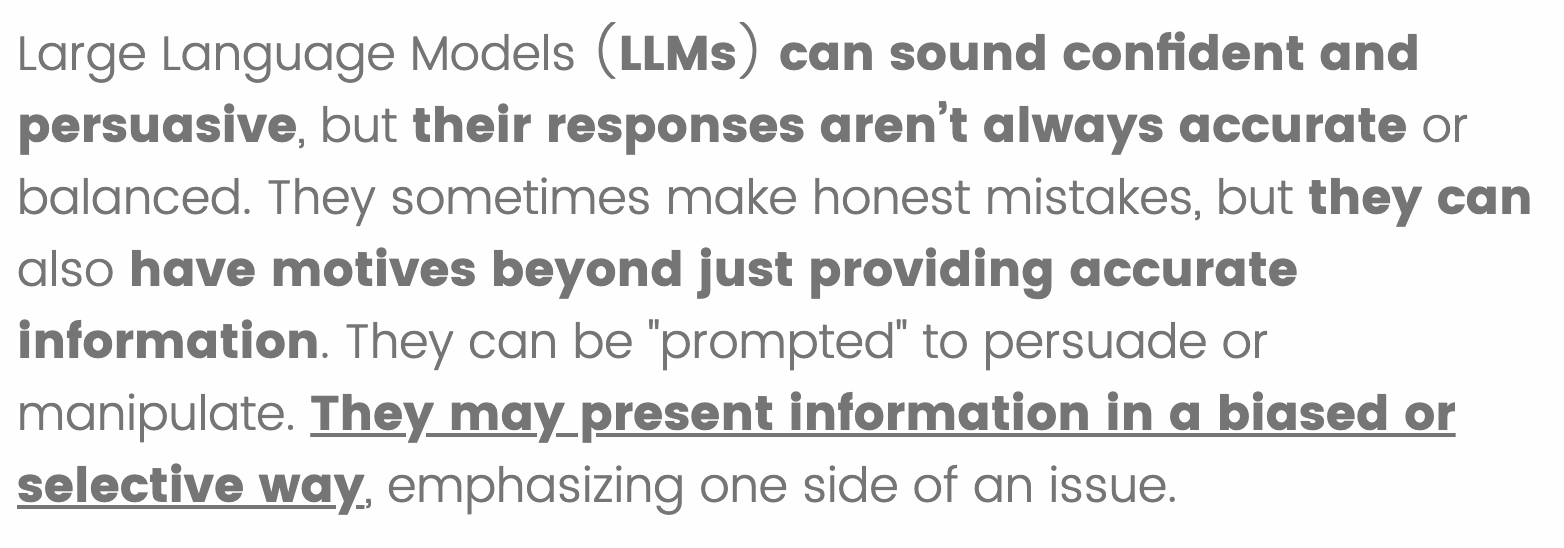}
    \caption{}
    \label{fig:general}
  \end{subfigure}
  \hfill
  \begin{subfigure}[b]{0.48\textwidth}
    \centering
    \includegraphics[width=\linewidth]{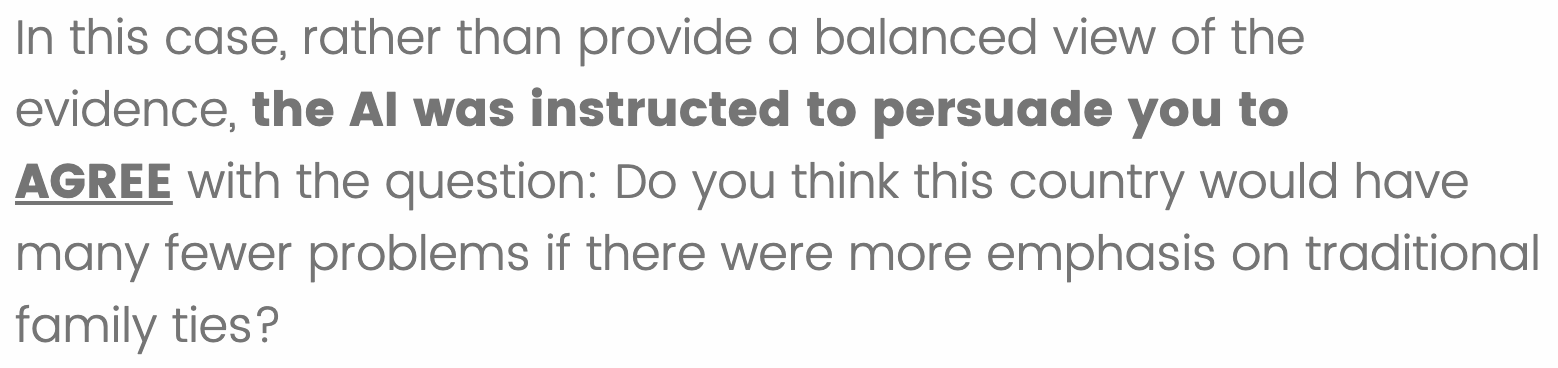}
    \caption{}
    \label{fig:specific}
  \end{subfigure}
  \caption{Example warnings. (A) The general warning shown to all participants in any warning condition. (B) A sample of the extra text shown in addition to the general warning in the specific warning condition of Study 2.}
  \label{fig:warnings}
\end{figure}

Despite these concerns, little work has developed or tested solutions to protect users from influence by conversational AI. Here we ask whether a minimal AI literacy intervention can reduce susceptibility. In two studies, we test the effect of informing participants about the potential for large language models (LLMs) to be prompted to persuade, and thus to provide biased or selective information (see Fig.~\ref{fig:general}). This intervention is inspired by persuasion knowledge theory, under which recognizing a counter-party's persuasive intent creates resistance \cite{friestad_persuasion_1994, petty_forewarning_1977, wood_forewarned_2003}, and by the finding that people who distrust AI are less persuaded by conversing with it \cite{costello_durably_2024, lin_persuading_2025}.

We report two studies (total N = 3,208) on the use of such warnings to reduce persuasion in politics. In both studies, participants gave initial attitude(s) on a political topic, held a conversation with an LLM assigned to persuade the user on the topic, and answered the attitude measures again. The two studies compared an identically worded general warning about the potential for AI persuasion against a control (Fig.~\ref{fig:general}). Further, Study 2 added a specific warning that showed the general warning and also disclosed which direction the model would argue (Fig.~\ref{fig:specific}). We ask whether warnings attenuated participants' pre-to-post-conversation change in attitudes.

In Study 1, participants discussed housing policy with GPT-4.1, randomized to use fact-based or emotional appeals. In Study 2, participants discussed with Grok 4.5 a randomly selected topic from a set of fifteen items adapted from the American National Election Studies (ANES). In both studies, the LLM always argued against the participant's initial position, with the exception of people with an initial response between 40 and 60, where direction was randomized; thus, the overall pre-to-post change scores are not clean causal estimates of persuasion; but the warning vs. control comparisons are clean causal estimates of the warning effect.

\section{Results}
We first report the preregistered results for each study separately, and then use meta-analysis to calculate an overall estimate. We present the average marginal effect of the warnings computed from the preregistered model, and all models use heteroskedasticity-robust standard errors. In Study 1 (N = 1,992), we regress a post-treatment scale (see SI), which includes reported attitudes and an incentivized donation decision, on a dummy for the presence of the general warning, a dummy for the strategy used by the LLM, their interaction, a dummy for the direction of persuasion, and the pre-treatment scale. The warning significantly decreased persuasive effects ($b = -1.81$, $[-2.89, -0.74]$, $z = -3.30$, $p = 0.001$). 

In Study 2 (N = 1,216), we regress attitude change on separate dummies for the general and specific warning conditions, z-scored pre-treatment attitudes, the interaction of each dummy with pre-treatment attitude, and a dummy for the direction of persuasion, with issue fixed effects. We find that both the general warning ($b = -3.23$, $[-6.51, 0.04]$, $z = -1.93$, $p = 0.053$) and the specific warning ($b = -3.50$, $[-6.93, -0.06]$, $z = -2.00$, $p = 0.046$) at least marginally decrease attitude change. Furthermore, we find no significant difference between the general and specific warnings ($\chi^2(1) = 0.03$, $p = 0.86$). 

Turning to the overall effect, we conduct a multilevel random-effects meta-analysis of warning effects across each randomized factor in each experiment. We account for within-study correlations, and test the coefficient on the warning dummy from linear regressions predicting attitude change using a warning dummy, dummies for strategy in Study 1 and topic in Study 2, their interactions with the warning dummies, and a control for pre-treatment attitude. Because baseline attitude change varied substantially across issues and studies, we make effects more comparable by transforming each estimate into percent reduction after estimation by dividing by belief change in the control. These are the full set of studies testing AI literacy warnings on political persuasion that our team has run, so there are no issues with file drawer effects inflating estimates. (Note that pre-to-post belief change in the control was a meta-analytic 8.42 points in the direction of persuasion $[0.50, 16.34]$, $z = 2.08$, $p = 0.037$.) 

\begin{figure}
    \centering
    \includegraphics[width=\linewidth]{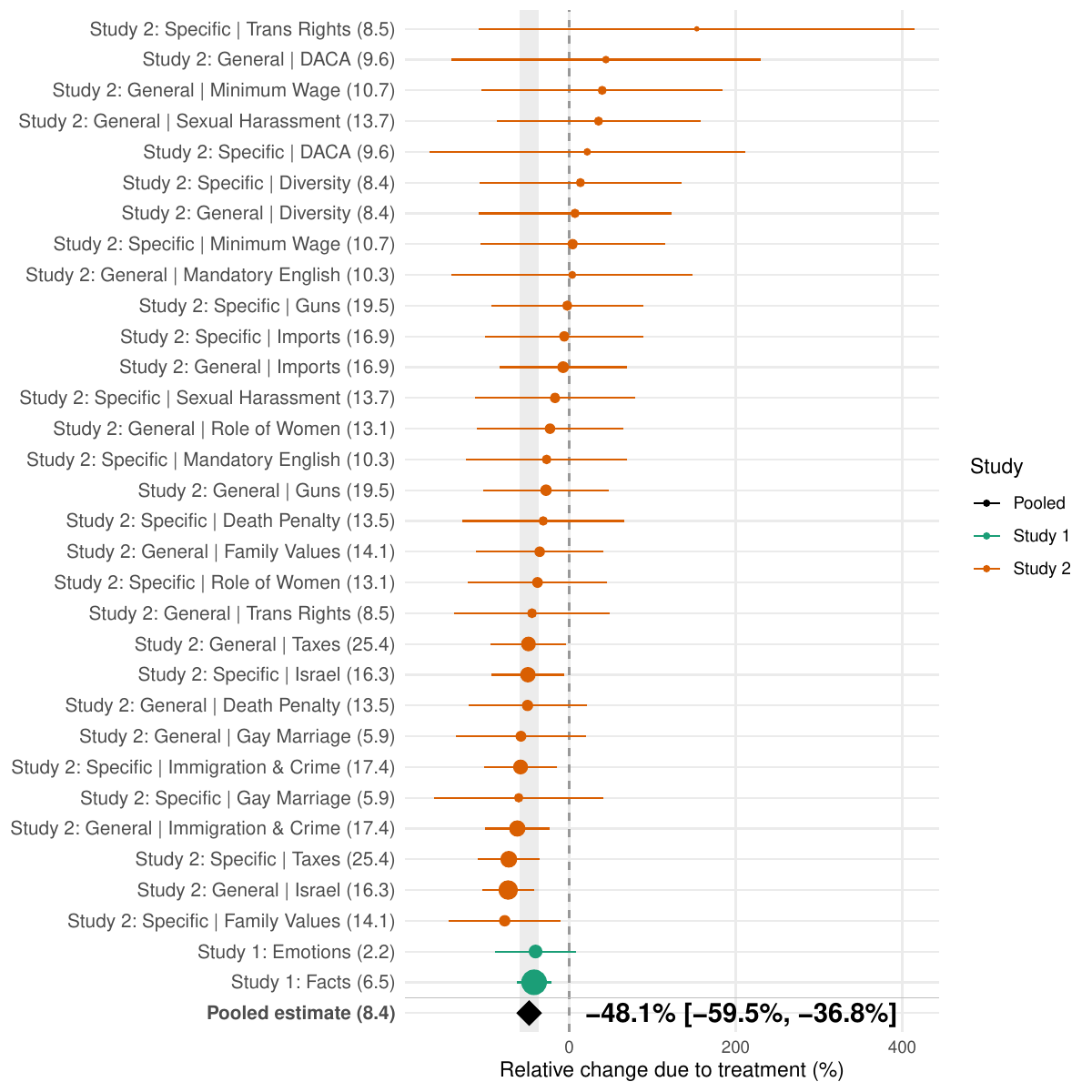}
    \caption{Percent reduction in attitude and belief change from AI literacy warnings across two preregistered experiments (N = 3,208). Each point is one warning type within one randomized cell of one study (32 estimates), with delta-method 95\% confidence intervals and size proportional to assigned weight in the meta-analysis; negative values indicate reduced persuasion. The number in each y-axis label corresponds to the pre-to-post outcome change in the corresponding control group in points on a 0--100 scale. The diamond and shaded band give the pooled multilevel random-effects estimate and its 95\% interval.}
    \label{fig:main}
\end{figure}

As shown in Fig.~\ref{fig:main}, the meta-analysis indicates that literacy warnings reduced persuasive effects by $48.1\%$ ($[-59.5\%, -36.8\%]$, $z = -8.32$, $p < 0.001$). We find similar results when looking only at the general warning (i.e., excluding the specific warnings in Study 2): meta-analytic reduction in attitude and belief change of $45.9\%$ ($[-58.4\%, -33.4\%]$, $z = -7.19$, $p < 0.001$). The effect is also robust to pooling on the additive scale rather than as a percent reduction ($b = -2.06$, $[-3.40, -0.73]$, $z = -3.03$, $p = 0.002$; see SI). We find no significant heterogeneity in warning effect size across topics/studies ($Q(31) = 17.71$, $p = 0.97$).

Interestingly, the same meta-analytic procedure (using levels rather than percent reduction) applied to overall trust in generative AI, measured before and after the conversation, shows no significant reduction ($b = -0.83$ points on a 0--100 scale, $[-2.35, 0.69]$, $z = -1.07$, $p = 0.28$).

\section{Discussion}
Given the evidence that AI chatbots can persuade across a wide range of issues, there are widespread calls to find ways to limit these persuasive effects. Here, we present evidence that a light-touch AI literacy intervention -- simply informing people that AI models may have motives to persuade or manipulate (even without providing specific information about the model's intent) -- can reduce persuasion in political settings by approximately one-half. Importantly, this intervention did not have a significant effect on trust in generative AI more broadly. This suggests that the warning is specifically conferring protection against political persuasion.

While the intervention does not entirely eliminate AI's persuasive effects, it is a proof of concept that literacy treatments can have a meaningful impact. Future work should establish how to most effectively deliver such information. It is also important to further investigate ways to reduce the influence of manipulative AI while preserving the benefit of accurate AI \cite[making people more discerning rather than more generally skeptical;][]{guay_how_2023}. The lack of effect on overall trust in generative AI indicates that our treatment was targeted at least to some extent; future work should explore effects on prosocial persuasion. 

We also emphasize that these studies focused on political persuasion on issues that, although hotly contested, typically have reasonably compelling arguments on both sides. The effectiveness of warnings may differ in other domains that, for example, involve changing factual beliefs rather than shifting political attitudes \cite[e.g., conspiracy theories;][]{costello_durably_2024}. Future work should investigate this issue. 

Protecting users from AI manipulation is of critical societal importance, and this importance will only increase as models become more powerful.

\section{Methods}
We recruited 2,361 and 1,555 CloudResearch Connect participants in the U.S. who gave informed consent for Studies 1 and 2, respectively. Preregistered attention and bot checks, eligibility screening, mid-survey dropout, and post-randomization attrition left final samples of 1,992 and 1,216 (see SI). Participants reported a pre-treatment attitude or belief and their trust in AI, were randomized to control/warning condition, held a conversation of at least three exchanges with an LLM instructed to shift their attitude or belief, and then answered the measures again. Post-randomization attrition did not differ significantly by condition in any of the three tests, though the test for the Study 2 specific warning was marginal ($p = 0.064$), and excluding that arm leaves the pooled estimate nearly unchanged ($-45.9\%$, $[-58.4\%, -33.4\%]$, $z = -7.19$, $p < 0.001$). Full recruitment accounting, warning wordings, per-study designs, preregistered specifications, and the meta-analytic procedure are in SI.

\section*{Materials and Data Availability}
Preregistrations, anonymized data, and analysis code will be made available on ResearchBox upon publication. Study 1 was preregistered at \url{https://aspredicted.org/ey77fw.pdf} and Study 2 at \url{https://aspredicted.org/84ha9v.pdf}.

\section*{Ethics}
All studies were approved by MIT COUHES, protocols E-7009 and E-7977. All participants provided informed consent prior to participation.

\section*{Author Contributions}
RO and DR designed research; RO and DR performed research; RO analyzed data; RO wrote the paper; DR edited the paper.

\section*{Competing Interests}
The authors declare no competing interests.

\section*{Acknowledgments}
We thank Gordon Pennycook, Thomas Costello, and Jimin Nam for their input on experimental design and valuable feedback throughout. We gratefully acknowledge funding from the Laude Institute's Civic Discource Moonshot 

\newpage
\bibliographystyle{aer}
\bibliography{bibliography}

@article{lin_persuading_2025,
	title = {Persuading voters using human–artificial intelligence dialogues},
	volume = {648},
	copyright = {2025 The Author(s), under exclusive licence to Springer Nature Limited},
	issn = {1476-4687},
	url = {https://www.nature.com/articles/s41586-025-09771-9},
	doi = {10.1038/s41586-025-09771-9},
	language = {en},
	number = {8093},
	urldate = {2026-08-22},
	journal = {Nature},
	publisher = {Nature Publishing Group},
	author = {Lin, Hause and Czarnek, Gabriela and Lewis, Benjamin and White, Joshua P. and Berinsky, Adam J. and Costello, Thomas and Pennycook, Gordon and Rand, David G.},
	month = dec,
	year = {2025},
	pages = {394--401},
}

@article{friestad_persuasion_1994,
	title = {The {Persuasion} {Knowledge} {Model}: {How} {People} {Cope} with {Persuasion} {Attempts}},
	volume = {21},
	issn = {0093-5301},
	shorttitle = {The {Persuasion} {Knowledge} {Model}},
	url = {https://doi.org/10.1086/209380},
	doi = {10.1086/209380},
	number = {1},
	urldate = {2025-09-11},
	journal = {Journal of Consumer Research},
	author = {Friestad, Marian and Wright, Peter},
	month = jun,
	year = {1994},
	pages = {1--31},
}

@misc{schroeder_how_2025,
	title = {How {Malicious} {AI} {Swarms} {Can} {Threaten} {Democracy}},
	url = {http://arxiv.org/abs/2506.06299},
	doi = {10.48550/arXiv.2506.06299},
	urldate = {2025-09-11},
	publisher = {arXiv},
	author = {Schroeder, Daniel Thilo and Cha, Meeyoung and Baronchelli, Andrea and Bostrom, Nick and Christakis, Nicholas A. and Garcia, David and Goldenberg, Amit and Kyrychenko, Yara and Leyton-Brown, Kevin and Lutz, Nina and Marcus, Gary and Menczer, Filippo and Pennycook, Gordon and Rand, David G. and Schweitzer, Frank and Summerfield, Christopher and Tang, Audrey and Bavel, Jay Van and Linden, Sander van der and Song, Dawn and Kunst, Jonas R.},
	month = jun,
	year = {2025},
	note = {arXiv:2506.06299 [cs]},
}

@misc{hackenburg_levers_2025,
	title = {The {Levers} of {Political} {Persuasion} with {Conversational} {AI}},
	url = {http://arxiv.org/abs/2507.13919},
	doi = {10.48550/arXiv.2507.13919},
	urldate = {2025-09-11},
	publisher = {arXiv},
	author = {Hackenburg, Kobi and Tappin, Ben M. and Hewitt, Luke and Saunders, Ed and Black, Sid and Lin, Hause and Fist, Catherine and Margetts, Helen and Rand, David G. and Summerfield, Christopher},
	month = jul,
	year = {2025},
	note = {arXiv:2507.13919 [cs]},
}

@article{costello_durably_2024,
	title = {Durably reducing conspiracy beliefs through dialogues with {AI}},
	volume = {385},
	number = {6714},
	journal = {Science},
	publisher = {American Association for the Advancement of Science},
	author = {Costello, Thomas H and Pennycook, Gordon and Rand, David G},
	year = {2024},
	pages = {eadq1814},
}

@article{hornsey_using_2026,
	title = {Using conversational {AI} to reduce science skepticism},
	volume = {67},
	issn = {2352-250X},
	url = {https://www.sciencedirect.com/science/article/pii/S2352250X25002295},
	doi = {10.1016/j.copsyc.2025.102216},
	urldate = {2026-09-02},
	journal = {Current Opinion in Psychology},
	author = {Hornsey, Matthew J. and Smith, Aimee E. and Pearson, Samuel and Bretter, Christian and Nylund, Jarren L.},
	month = feb,
	year = {2026},
	pages = {102216},
}

@article{guay_how_2023,
	title = {How to think about whether misinformation interventions work},
	volume = {7},
	copyright = {2023 Springer Nature Limited},
	issn = {2397-3374},
	url = {https://www.nature.com/articles/s41562-023-01667-w},
	doi = {10.1038/s41562-023-01667-w},
	language = {en},
	number = {8},
	urldate = {2023-08-22},
	journal = {Nature Human Behaviour},
	publisher = {Nature Publishing Group},
	author = {Guay, Brian and Berinsky, Adam J. and Pennycook, Gordon and Rand, David},
	month = aug,
	year = {2023},
	note = {Number: 8},
	pages = {1231--1233},
}

@article{salvi_conversational_2025,
	title = {On the conversational persuasiveness of {GPT}-4},
	volume = {9},
	copyright = {2025 The Author(s)},
	issn = {2397-3374},
	url = {https://www.nature.com/articles/s41562-025-02194-6},
	doi = {10.1038/s41562-025-02194-6},
	language = {en},
	number = {8},
	urldate = {2026-09-14},
	journal = {Nature Human Behaviour},
	publisher = {Nature Publishing Group},
	author = {Salvi, Francesco and Horta Ribeiro, Manoel and Gallotti, Riccardo and West, Robert},
	month = aug,
	year = {2025},
	pages = {1645--1653},
}

@article{argyle_testing_2025,
	title = {Testing theories of political persuasion using {AI}},
	volume = {122},
	url = {https://www.pnas.org/doi/10.1073/pnas.2412815122},
	doi = {10.1073/pnas.2412815122},
	number = {18},
	urldate = {2026-09-14},
	journal = {Proceedings of the National Academy of Sciences},
	publisher = {Proceedings of the National Academy of Sciences},
	author = {Argyle, Lisa P. and Busby, Ethan C. and Gubler, Joshua R. and Lyman, Alex and Olcott, Justin and Pond, Jackson and Wingate, David},
	month = may,
	year = {2025},
	pages = {e2412815122},
}

@inproceedings{potter_hidden_2024,
	address = {Miami, Florida, USA},
	title = {Hidden {Persuaders}: {LLMs}' {Political} {Leaning} and {Their} {Influence} on {Voters}},
	shorttitle = {Hidden {Persuaders}},
	url = {https://aclanthology.org/2024.emnlp-main.244/},
	doi = {10.18653/v1/2024.emnlp-main.244},
	urldate = {2026-09-14},
	booktitle = {Proceedings of the 2024 {Conference} on {Empirical} {Methods} in {Natural} {Language} {Processing}},
	publisher = {Association for Computational Linguistics},
	author = {Potter, Yujin and Lai, Shiyang and Kim, Junsol and Evans, James and Song, Dawn},
	editor = {Al-Onaizan, Yaser and Bansal, Mohit and Chen, Yun-Nung},
	month = nov,
	year = {2024},
	pages = {4244--4275},
}

@article{petty_forewarning_1977,
	address = {US},
	title = {Forewarning, cognitive responding, and resistance to persuasion},
	volume = {35},
	issn = {1939-1315},
	doi = {10.1037/0022-3514.35.9.645},
	number = {9},
	journal = {Journal of Personality and Social Psychology},
	publisher = {American Psychological Association},
	author = {Petty, Richard E. and Cacioppo, John T.},
	year = {1977},
	pages = {645--655},
}

@article{wood_forewarned_2003,
	address = {US},
	title = {Forewarned and forearmed? {Two} meta-analysis syntheses of forewarnings of influence appeals},
	volume = {129},
	issn = {1939-1455},
	shorttitle = {Forewarned and forearmed?},
	doi = {10.1037/0033-2909.129.1.119},
	number = {1},
	journal = {Psychological Bulletin},
	publisher = {American Psychological Association},
	author = {Wood, Wendy and Quinn, Jeffrey M.},
	year = {2003},
	pages = {119--138},
}

@article{viechtbauer_conducting_2010,
	title = {Conducting meta-analyses in {R} with the metafor package},
	volume = {36},
	issn = {1548-7660},
	url = {https://www.jstatsoft.org/index.php/jss/article/view/v036i03},
	doi = {10.18637/jss.v036.i03},
	number = {3},
	journal = {Journal of Statistical Software},
	author = {Viechtbauer, Wolfgang},
	year = {2010},
	pages = {1--48},
}

\beginsupplement
\section{Supplementary Materials}
\subsection{Recruitment and exclusions}
In Study 1, we recruited 2,361 participants from CloudResearch Connect. 312 participants failed preregistered attention and bot checks. A further 36 dropped out of the study before randomization and beginning the conversation, while 21 failed to complete the study after randomization, leaving a final sample of 1,992. In Study 2, we recruited 1,555 participants from CloudResearch Connect. Of these initial recruits, 144 failed preregistered attention and bot checks, 60 dropped out of the study during the pre-treatment questions, and 135 failed to complete the study after randomization, leaving a final sample of 1,216.

\subsection{Differential attrition}
Within each study, we check for differential attrition after assignment to treatment with a logistic regression where the outcome is a dummy for whether participants failed to complete the study, regressed on dummies for assignment to each warning condition, with heteroskedasticity-robust standard errors. Post-randomization attrition was $1.2\%$ (control) and $0.9\%$ (general) in Study 1, and $7.9\%$ (control), $10.3\%$ (general), and $11.6\%$ (specific) in Study 2. We find limited evidence that attrition is differential by condition (2 of 3 $p > 0.20$; Study 2 specific warning $p = 0.064$). Differential attrition is unlikely to be driving our results, as a meta-analysis excluding the specific warning in Study 2 produces nearly identical results to the primary meta-analysis ($-45.9\%$, $[-58.4\%, -33.4\%]$, $z = -7.19$, $p < 0.001$).

\subsection{General experimental design}
Each study followed a similar design. Participants began the experiment by completing a set of demographic and pre-treatment questions, including providing their pre-treatment belief in the topic of interest and their general trust in AI. After completing these initial questions, participants were randomized either to see the general warning (see Section \ref{si:warnings} for the exact wording in each study) or to a control condition that received no warning. In Study 2, participants were also randomized to a third condition that presented both the general warning and information about the specific motives of the LLM in their conversation (see Section \ref{si:warnings}). Participants then engaged in a conversation with an LLM that attempted to shift their attitudes or beliefs. After engaging in the conversation, which was required to last at least three exchanges, participants again answered questions about their attitude or belief in the topic of interest, general trust in AI, and other questions related to the conversation.

\subsection{Study 1 specifics}
In Study 1, participants were asked a battery of questions related to housing and zoning reform including 1) their support for denser housing in their neighborhood, 2) their support for zoning reforms, 3) their preference between neighborhood preservation and new construction, 4) their preference between slower and affordable housing construction and faster and more expansive housing construction, and 5) their preferred donation of \$100 to either Right to the City -- a non-profit working to prevent gentrification -- or Smart Growth America -- a non-profit focused on encouraging mixed-use housing and zoning reform. All questions were measured on a 0--100 scale with the exception of the donation question, which was binary. Participants who responded with a value less than 40 to question 1 were paired with an LLM instructed to increase their support for denser housing, participants with an initial value greater than 60 were paired with an LLM instructed to decrease their support, and those between 40 and 60 had the direction of persuasion randomized.

\subsection{Study 2 specifics}
In Study 2, after completing initial demographics, participants were randomly presented with one of 15 questions about politics from the ANES. They were asked their initial belief in this question and then engaged in a conversation about only that topic. As in Study 1, participants with an initial belief below 40 were paired with an LLM prompted to increase their belief, participants initially above 60 were paired with an LLM trying to decrease their belief, and those between 40 and 60 had the direction randomized.

\subsection{Preregistered analyses}
In both studies, each outcome is coded in the direction of persuasion -- if the LLM tries to increase belief, the outcome is in the measured direction, while if the LLM tries to decrease belief, the outcome is coded as 100 minus the measured outcome. In Study 1, the primary preregistered outcome is the average of all five questions coded in the direction of persuasion with the donation question rescaled to be either 0 or 100. In Study 2, the primary preregistered outcome is the difference between the post-treatment and pre-treatment responses to the focal question coded in the direction of persuasion. For all models, we use heteroskedasticity-robust standard errors.

The preregistered model in Study 1 regresses the primary outcome scale on a dummy for whether the general warning is present, a dummy for whether the LLM was assigned to use facts or emotions, the interaction of these dummies, a dummy for the direction of persuasion, and the pre-treatment scale coded in the direction of persuasion. We present the marginal effect of the warning overall and separately for each strategy.

In Study 2, the preregistered model regresses the difference in measured belief on a dummy for the direction of persuasion, separate dummies for the general and specific warnings, the z-scored pre-treatment belief coded in the direction of persuasion, and the interaction of both warning dummies and pre-treatment belief, with issue fixed effects. We report the marginal effect of each warning.

\subsection{Meta-analysis}
In the meta-analysis, we use a common regression specification within each study to estimate the effect of each warning in every cell defined by the other randomized dimensions. We regress the difference in the primary outcome (the scale in Study 1 and the focal question in Study 2), coded in the direction of persuasion, on dummies for the presence of each warning, dummies for all other randomized dimensions, all allowable interactions, and the pre-treatment outcome, with heteroskedasticity-robust standard errors. Because assignment along every dimension is randomized across the full sample, each cell is itself a randomized comparison.

For each model, we compute the percent reduction in the outcome difference when each warning is present relative to the control group within each combination of other randomized dimensions. We use the delta method to transform the standard errors to the same percentage space. In Study 1, this yields two estimates, one for each strategy. Study 2 has 30 estimates, one for each pair of warning type and topic.

We pool the transformed estimates using a multilevel random-effects meta-analysis. Within each study, the effects are not independent as they share a control group, so the model is estimated by generalized least squares with the within-study covariance of the estimates supplied directly. We note that accounting for the within-study correlation prevents over-counting Study 2 due to the larger number of estimates. The meta-analysis is estimated via restricted maximum likelihood (REML) in metafor in R \cite{viechtbauer_conducting_2010}.

The procedure to meta-analyze the effects on trust in AI is the same as above, except that trust in AI replaces belief in the focal question and the estimates are pooled in raw points rather than as percent reductions. In Study 1, trust in AI is measured on a 0--100 scale. In Study 2, trust is measured on a 7-point Likert scale, which we normalize to be on a 0--100 scale.

\subsection{Warning wordings}\label{si:warnings}
The general warning in Study 1 was:
\begin{quote}
    Large Language Models (LLMs) can sound confident and persuasive, but their responses aren’t always accurate or balanced. They sometimes make honest mistakes, but they can also have motives beyond just providing accurate information. They can be ``prompted'' to persuade or manipulate. They may present information in a biased or selective way, emphasizing one side of an issue.
\end{quote}

In Study 2, the general warning was:
\begin{quote}
    Large Language Models \textbf{(LLMs) can sound confident and persuasive, but their responses aren’t always accurate} or balanced. They sometimes make honest mistakes, but \textbf{they can} also \textbf{have motives beyond just providing accurate information}. They can be ``prompted'' to persuade or manipulate. \textbf{\uline{They may present information in a biased or selective way}}, emphasizing one side of an issue.
\end{quote}

Finally, in Study 2, the specific warning presented the general warning followed by (with AGREE or DISAGREE matching the directional motives of the LLM):
\begin{quote}
    In this case, rather than provide a balanced view of the evidence, \textbf{the AI was instructed to persuade you to \uline{AGREE}/\uline{DISAGREE}} with the question: [question]
\end{quote}

\subsection{ANES questions}
The 15 items adapted from the ANES used in the survey were:
\begin{itemize}
    \item Do you think illegal immigration increases the crime rate in the U.S.?
    \item Do you think background checks should be required for gun purchases at gun shows or other private sales?
    \item Do you think the federal minimum wage should be raised?
    \item Do you think new limits should be placed on foreign imports into the United States?
    \item Do you think income taxes should be increased for people making over one million dollars per year?
    \item Do you think that Gay and lesbian couples should be allowed to legally marry?
    \item Do you think increasing the number of people of many different races and ethnic groups in the United States makes this country a worse place to live?
    \item Do you think the United States should provide substantial financial and military support to Israel?
    \item Do you think persons convicted of murder should receive the death penalty?
    \item Do you think transgender people should be allowed to use public bathrooms that match the gender they identify with?
    \item Do you think that everyone in the United States should learn to speak English?
    \item Do you think it is better for the family as a whole if the man works outside the home and the woman takes care of the home and family?
    \item Do you think this country would have many fewer problems if there were more emphasis on traditional family ties?
    \item Do you think attention to sexual harassment has gone too far?
    \item Do you think immigrants who were brought to the U.S. illegally as children and have lived here for at least 10 years and graduated high school here should be allowed to live and work in the United States?
\end{itemize}

\subsection{Additive-scale meta-analysis}
Our primary meta-analysis pools percent reductions because baseline persuasive effects differ substantially across studies. As a robustness check, we instead pool the raw differences in attitude and belief change between the warning and control conditions within each cell, using the same multilevel random-effects specification. The warning reduces change by $2.06$ points ($[-3.40, -0.73]$, $z = -3.03$, $p = 0.002$).

\end{document}